\documentclass[prl,twocolumn,superscriptaddress,showpacs,a4paper]{revtex4}

\usepackage{times} \normalfont \usepackage[T1]{fontenc}
\usepackage{graphicx}

\def\xx{\hbox{\sffamily\bfseries x}}
\def\yy{\hbox{\sffamily\bfseries y}}

\def\uu{\hbox{\sffamily\bfseries u}}
\def\vv{\hbox{\sffamily\bfseries v}}
\def\aa{\hbox{\sffamily\bfseries a}}
\def\ss{\hbox{\sffamily\bfseries s}}
\def\QQ{\hbox{\sffamily\bfseries Q}}
\def\PP{\hbox{\sffamily\bfseries P}}

\begin{document}

\title{
  A new and efficient approach to time-dependent \\ density-functional
  perturbation theory for optical spectroscopy
}

\author{
  Brent Walker
} 
\email{
  bwalker@ictp.it
} 
\affiliation{
  ICTP -- The Abdus Salam International Centre for Theoretical Physics,
  Strada Costiera 11, I-34014 Trieste, Italy
}
\affiliation{
  CNR-INFM, DEMOCRITOS National Simulation Center, Trieste, Italy
}

\author{
  A. Marco Saitta
}
\affiliation{
 Physique des Milieux Condens\'es, Universit\'e Pierre et Marie Curie,
  Paris, France
}

\author{
  Ralph Gebauer
} 
\affiliation{
  ICTP -- The Abdus Salam International Centre for Theoretical Physics,
  Strada Costiera 11, I-34014 Trieste, Italy
}
\affiliation{
  INFM -- DEMOCRITOS National Simulation Center, Trieste, Italy
}

\author{
  Stefano Baroni
} 
\email{
  baroni@sissa.it
}
\affiliation{
  SISSA -- Scuola Internazionale Superiore di Studi Avanzati, Via
  Beirut 2-4, I-34014 Trieste, Italy
}
\affiliation{
  INFM -- DEMOCRITOS National Simulation Center, Trieste, Italy
}

\date{\today}

\begin{abstract}
  Using a super-operator formulation of linearized time-dependent
  density-functional theory, the dynamical polarizability of a system
  of interacting electrons is given a matrix continued-fraction
  representation whose coefficients can be obtained from the
  non-symmetric block-Lanczos method. The resulting algorithm allows
  for the calculation of the {\em full spectrum} of a system with a
  computational workload which is only a few times larger than that
  needed for {\em static} polarizabilities within time-independent
  density-functional perturbation theory. The method is demonstrated
  with the calculation of the spectrum of benzene, and prospects
  for its application to the large-scale calculation of optical
  spectra are discussed.
\end{abstract}


\maketitle

The past two decades have witnessed the tremendous success of
density-functional theory (DFT) \cite{HK,KS} in describing and
predicting various properties of systems of interacting electrons,
ranging from atoms and molecules to solids and liquids. In its
original formulation, DFT only applies to the electronic ground
state. This limitation was lifted by Runge and Gross (RG) \cite{RG}
who generalized DFT to time-dependent systems. According to the RG
theorem, for any given ($t=0$) initial state of the many-body system,
the external, generally time-dependent, potential acting on it is
uniquely determined by the time evolution of the one-electron density,
$n(\mathbf{r},t)$, for $t>0$. Based on this theorem, it is possible to
formally establish a time-dependent Kohn-Sham (KS) equation from which
various one-particle properties of the system can be obtained as
functions of time. The resulting theoretical framework is usually
referred to as {\em time-dependent density-functional theory}
(TDDFT). Unfortunately, little is known about the exchange-correlation
(XC) potential in ordinary DFT, and even less is known about it in the
time-dependent case. Most of the existing applications of TDDFT are
based on the {\em adiabatic local density} (ALDA) or {\em adiabatic
  generalized gradient} approximations, which amount to using the same
functional dependence of the XC potential upon density as in the
static case. In spite of the crudeness of these approximations, the
optical spectra calculated using them is in some cases almost as
accurate as those obtained from more computationally demanding
many-body approaches \cite{Onida}. TDDFT is in principle an exact
theory. Progress in understanding and characterizing the XC functional
will thus substantially increase the predictive power of TDDFT while
keeping its computational requirements at a significantly more modest
level than methods based on many-body perturbation theory.

Linearization of TDDFT with respect to the strength of some external
perturbation to an otherwise time-independent system leads to a
non-Hermitean eigenvalue problem whose eigenvalues are excitation
energies of the system, and whose eigenvectors can be used to
calculate appropriate oscillator strengths \cite{Casida}. Not
surprisingly, this eigenvalue problem has the same structure as that
arising from time-dependent Hartree-Fock theory \cite{TDHF}, and the
dimension of the resulting matrix (the {\em Liouvillian}) is thus
twice the product of the number of occupied states, $N_{v}$, with the
number of empty states. The diagonalization of such a large matrix can
be accomplished using iterative techniques, often in conjunction with
the so-called Tamm-Dancoff approximation (TDA) which amounts to
enforcing Hermiticity by neglecting the anti-Hermitean component of
the Liouvillian \cite{Tamm-Dancoff}. Many of the existing molecular
applications of TDDFT have been performed within such a framework
which suffers however from two major limitations. The first is that
many individual eigenvalues/eigenvectors must be calculated in order
to model the spectral properties in any given frequency range, and the
number of such eigenpairs increases with the size of the system. The
second limitation is that a straightforward implementation of this
method preliminarily requires the full diagonalization of the
unperturbed Kohn-Sham Hamiltonian, a task which may become prohibitive
for large systems and/or when the one-electron basis set is very large
(as may be the case with plane waves or real-space grids). The first
of these problems is sometimes dealt with by directly calculating the
relevant response function(s), rather than individual excitations
\cite{Onida}. The price paid in this case is the manipulation
(inversion, multiplication) of large matrices for any individual
frequency, a task which may again be impossible for large
systems/basis sets. Even so, the full diagonalization of the
unperturbed KS Hamiltonian cannot be avoided.

An alternative approach to TDDFT, first proposed by Yabana and Bertsch
\cite{YB}, avoids diagonalization altogether. In this approach, the
TDDFT KS equations are solved in the time domain and susceptibilities
are obtained by Fourier analyzing the response of the system to
appropriate perturbations in the linear regime. This scheme has the
same computational complexity as standard time-independent iterative
methods in DFT. For this reason, real-time methods have recently
gained popularity in conjunction with the use of real-space grids
\cite{AR}, and a similar success should be expected using plane-wave
(PW) basis sets. The main limitation of these methods is that the time
step needed for a stable integration of the TDDFT KS equations is
small (of the order of $10^{-3}$ fs in typical pseudo-potential
applications) and decreases as the number of PW's (or real-space grid
points) increases.

In this Letter we propose a novel way to calculate optical spectra in
the frequency domain---thus avoiding any explicit integration of the
TDDFT KS equations---which does not require any diagonalization (of
either the unperturbed KS Hamiltonian, or the TDDFT Liouvillian), nor
any time-consuming matrix operations. To this end, we first express a
generalized susceptibility as a matrix element of the resolvent of the
Liouvillian super-operator, defined in some appropriate operator
space, between suitable operator states. This matrix element is then
evaluated using a Lanczos continued-fraction technique. Each link of
the Lanczos chain---that is calculated once for all
frequencies---requires a number of floating-point operations which is
only a few times larger than that needed by a single step of the
iterative calculation of a {\em static} polarizability within
time-independent density-functional perturbation theory (DFPT)
\cite{BGT,RMP}.

Let us first consider an external perturbation whose Fourier transform
will be indicated by $\lambda(\omega) \tilde V{'}_{ext}
(\mathbf{r},\omega)$, where $\lambda(\omega)$ conventionally indicates
the strength of the perturbation. Let $\varphi_{v}^{\circ}
(\mathbf{r})$ and $\epsilon_{v}$ be the ground-state valence KS
orbitals and energies, and $\varphi_{v} (\mathbf{r},t)$ the solution
of the TDDFT KS equation with the initial condition
$\varphi_{v}(\mathbf{r},0) = \varphi_{v}^\circ(\mathbf{r})$. We define
the orbital response functions as $ \varphi_{v}'(\mathbf{r},t) =
\mathrm{e}^{i\epsilon_{v}t} \bigl ( \varphi_{v} (\mathbf{r},t) -
\varphi_{v}^{\circ} (\mathbf{r}) \bigr )$, and indicate their Fourier
transforms as $\varphi^{+}_{v}(\mathbf{r}) = \tilde
\varphi'_{v}(\mathbf{r},\omega)$, and $\varphi^{-}_{v}(\mathbf{r}) =
\tilde \varphi'^{*}_{v}(\mathbf{r},-\omega)$. The $\varphi^{\pm}_{v}$
orbitals can be chosen to be orthogonal to the Kohn-Sham occupied
manifold, and they can be easily shown to satisfy the coupled linear
equations:
\begin{eqnarray}
  \phantom{-}\omega \varphi^{+}_{v}(\mathbf{r}) &=& \left(\hat{H}_{KS}^{\circ} -
    \epsilon_{v}\right) \varphi^{+}_{v}(\mathbf{r}) +  \hat P_{c}\tilde
  V'(\mathbf{r},\omega)\varphi_{v}^{\circ} (\mathbf{r}), \nonumber \\ 
  -\omega \varphi^{-}_{v}(\mathbf{r}) &=& \left(\hat{H}_{KS}^{\circ} -
    \epsilon_{v} \right) \varphi^{-}_{v}(\mathbf{r}) + \hat P_{c} \tilde
    V'(\mathbf{r},\omega) \varphi_{v}^{\circ} (\mathbf{r}), \hbox{\hglue
    5pt} \label{eq:KS-omega}
\end{eqnarray}
where $\hat{H}^{\circ}_{KS}$ is the ground-state KS Hamiltonian, $\hat
P_{c}$ is the projector onto the KS empty-state manifold, $ \tilde
V'(\mathbf{r},\omega) = \tilde V'_{ext} (\mathbf{r},\omega) + \int
\kappa (\mathbf{r},\mathbf{r}') \tilde n'(\mathbf{r}',\omega)
d\mathbf{r}'$ is the perturbing KS potential, $
\tilde{n}'(\mathbf{r},\omega ) = \sum_{v=1}^{N_v}\varphi_{v}^\circ
(\mathbf{r}) \bigl(\varphi^{+}_{v}(\mathbf{r}) +
\varphi^{-}_{v}(\mathbf{r})\bigr)$ is the linear variation of the
electron density, and $ \kappa (\mathbf{r},\mathbf{r}') =
\frac{e^{2}}{|\mathbf{r}-\mathbf{r}'|} + \left .  \frac{\delta
    V_{XC}(\mathbf{r})}{\delta n(\mathbf{r}')} \right |_{n=
  n^{\circ}}$ is the Hartree plus XC kernel. In
Eq.~(\ref{eq:KS-omega}) use has been made of the fact that the
perturbation is real $[\tilde V'^{*}(\mathbf{r},-\omega) =
\tilde V'(\mathbf{r},\omega) ] $, while in the expression for
$\tilde n'$, the $\varphi^{\circ}$'s are assumed to be real due to
time-reversal invariance. Following Refs.~\onlinecite{hutter} and
\onlinecite{tsiper}, we now consider the {\em super-vector} $|X\rangle
\equiv | \xx , \yy \rangle $, where \xx\ and \yy\ are {\em batches} of
orbitals, $\xx=\{x_v(\mathbf{r})\}$, $\yy=\{y_v(\mathbf{r})\}$,
defined as: $ x_{v}(\mathbf{r}) = {1\over 2} [
\varphi^{+}_{v}(\mathbf{r}) + \varphi^{-}_{v}(\mathbf{r}) ] $, and
$ y_{v}(\mathbf{r}) = {1\over 2} [ \varphi^{+}_{v}(\mathbf{r})
- \varphi^{-}_{v}(\mathbf{r})  ] $. 
The fundamental TDDFT equations, Eq.~(\ref{eq:KS-omega}), can then be
formally written as: 
\begin{equation}
  (\omega - {\cal L} ) |\xx,\yy \rangle = |0,\vv \rangle,
  \label{eq:liouville} 
\end{equation}
where the batch $\vv$ is defined as $ \vv = \{ \hat P_{c} \tilde
V'_{ext}(\mathbf{r},\omega) \varphi^{\circ}_{v}(\mathbf{r}) \}$, the
Liouvillian {\em super-operator} $\mathcal{L}$ is defined as: $
\mathcal{L} |\xx,\yy\rangle = |\mathcal{D}\cdot \yy,(\mathcal{D} +
\mathcal{K}) \cdot \xx \rangle$,
and $\mathcal{D}$ and $\mathcal{K}$ are Hermitean operators acting on
batches of response orbitals, $\uu = \{u_{v}(\mathbf{r}) \}$, as: 
\begin{eqnarray}
  \mathcal{D} \cdot \uu  &=& \{ ( \hat{H}^{\circ}_{KS} -
  \epsilon_{v} )u_{v}(\mathbf{r}) \}, \label{eq:D} \\ 
  \mathcal{K} \cdot \uu &=& \left \{ 
    \varphi^{\circ}_{v}(\mathbf{r}) \sum_{v'} \int
    \kappa(\mathbf{r},\mathbf{r'}) 
    \varphi^{\circ}_{v'}(\mathbf{r}') u_{v'}(\mathbf{r}')
    d\mathbf{r'} \right 
  \}. \label{eq:K}
\end{eqnarray}
Eq.~(\ref{eq:liouville}) gives the response of the system to the
external perturbation, as a function of frequency. When $V'_{ext}=0$,
this equation reduces to a (non-Hermitean) eigenvalue problem whose
eigenvectors describe the free oscillations of the system
corresponding to electronic excitations. This is essentially
equivalent to Casida's formulation of TDDFT \cite{Casida}.

In practice, one is seldom interested in the response of the system to
the most general perturbation, or in individual excitation energies,
but just in the frequency-dependent response of some {\em specific}
property to some {\em specific} perturbation. Let us consider an
observable, $\hat A$, whose time-dependent linear response is given
by: $ A(t) = 2 \mathrm{Re} \sum_{v} \langle \varphi^{\circ}_{v} |
\varphi'_{v}(t) \rangle $. Assuming that $\hat A$ is time-reversal
invariant, the Fourier transform of $A(t)$ can be written as: $ \tilde
A(\omega) = \sum_{v} ( \langle \varphi^{\circ}_{v} | \hat A |
\varphi^{+}_{v} \rangle + \langle \varphi^{\circ}_{v} | \hat A |
\varphi^{-}_{v} \rangle ) \equiv 2 \langle \aa,0|\xx,\yy \rangle $,
where $ \aa = \{ \hat A \varphi^{\circ}_{v}\}$. A generalized
susceptibility, $\chi_{AV}(\omega)$, can be defined as the derivative of
$\tilde A(\omega)$ with respect to the strength of the perturbation,
$\lambda(\omega)$. Using Eq.~(\ref{eq:liouville}) and the linearity of
$\tilde A(\omega)$ with respect to $\lambda(\omega)$, the
susceptibility can be written as:
\begin{equation}
  \chi_{AV}(\omega) = 2 \langle \aa,0 | (\omega - {\cal L})^{-1} |
  0,\vv \rangle . \label{eq:susceptibility}
\end{equation}
The results obtained so far and embodied in
Eq.~(\ref{eq:susceptibility}) can be summarized by saying that within TDDFT
any generalized susceptibility can be expressed as an appropriate {\em
  off-diagonal} matrix element of the resolvent of the Liouvillian
super-operator. In the following we show how such a matrix element can
be conveniently calculated using a matrix continued fraction resulting
from the non-Hermitean block-Lanczos algorithm (NHBLA) \cite{Bai-and-Day}.

Let us define a {\em block}, $|\QQ\rangle $, as a pair of orthogonal
super-vectors: $|\QQ\rangle= \{ |Q_1 \rangle, |Q_2\rangle \}$. The
scalar product between two blocks, $\ss = \langle \PP | \QQ \rangle$
is defined as the $2\times 2$ matrix: $s_{ij} = \langle P_i | Q_j
\rangle$, and the action of a super-operator on a block is defined as
the block whose elements are the result of the action of the
super-operator on each of the two elements of the original block:
${\cal L}|\{Q_1,Q_2\}\rangle \equiv |\{ {\cal L}Q_1, {\cal L}Q_2 \}
\rangle$. Given two starting blocks, $| \QQ^1 \rangle $ and $| \PP^1
\rangle $ such that $\langle P^1_i | Q^1_j \rangle = \delta_{ij}$, the
NHBLA allows one to generate a sequence of block pairs, $\{ | \QQ^n
\rangle, | \PP^n \rangle \}$, such that: $\langle P^n_i | Q^m_j
\rangle = \delta_{mn} \delta_{ij} $, and $ {\cal L} = \sum_{nm,ij}
T^{mn}_{ij} |Q^m_i\rangle\langle P^n_j |$, where: $ \quad T^{nm}_{ij}
= \langle P^n_i | {\cal L} | Q^m_j \rangle = a^n_{ij} \delta_{mn} +
b^{n}_{ij} \delta_{m+1,n} + c^{m}_{ij} \delta_{m,n+1} $ is a
block-tridiagonal matrix. Using these relations, the projection of the
resolvent of the Liouvillian over the starting block pair can then be
easily expressed as a matrix continued fraction:
\begin{equation}
  \langle \PP_1 | (\omega - {\cal L})^{-1} | \QQ_1 \rangle =
  \frac{{\displaystyle 1}}{{\displaystyle \omega - {\bf a}_1 + {\bf
        b}_2  \frac{{\displaystyle 1}}{{\displaystyle \omega - {\bf
            a}_2 + \cdots } } } {\bf c}_2},  
  \label{eq:continued-fraction}
\end{equation}
where the $\mathbf{a}$'s, $\mathbf{b}$'s, $\mathbf{c}$'s, are $2
\times 2$ matrices. If the starting block is chosen as $|Q^1_1\rangle
= |\aa,0 \rangle$, and $|Q^1_2\rangle = |0, \vv \rangle$, the
generalized susceptibility, Eq.~(\ref{eq:susceptibility}), is then the
$(1,2)$ matrix element of the $2\times 2$ matrix given by
Eq.~(\ref{eq:continued-fraction}). Without going into the details of
the NHBLA, suffice it to say that its implementation does not require
the explicit calculation of the Liouvillian super-operator, nor even
of the unperturbed KS Hamiltonian, but just the availability of a
black-box computer routine which, for any given batch of response
functions, $\uu = \{ u_v({\bf r}) \}$, returns ${\cal D}|\uu \rangle$
and ${\cal K} | \uu \rangle$, according to Eqs.~(\ref{eq:D}) and
(\ref{eq:K}). Each step of the NHBLA essentially involves two calls to
such a routine whose computational cost is roughly the same as that of
a single iteration in a static DFPT calculation.

\begin{figure}[!bt]
  \begin{center}
    \includegraphics[width=80mm]{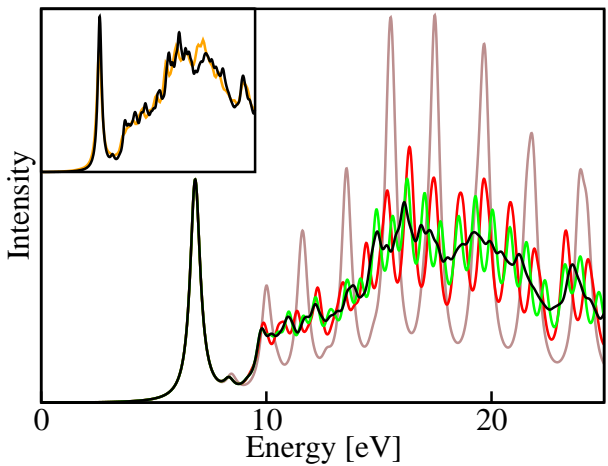}
    \caption{
      \label{fig:benzene_full_int_convergence} 
      Absorption spectrum of benzene calculated using the Lanczos
      method with different numbers of recursion steps: 1000 (plum),
      2000 (red), 3000 (green), and 6000 (black). The inset compares
      the 6000-step spectrum (black) with that obtained using the
      real-time propagation method (orange).  }
  \end{center}
\end{figure}

The theory described above was implemented on top of the {\tt PWscf}
plane-wave pseudopotential code which is part of the {\em Quantum
  ESPRESSO} distribution \cite{espresso}. As a test to demonstrate our
methodology, we have calculated the absorption spectrum of benzene, a
system for which many excited-state calculations already exist
\cite{benzene-theory,benzene-tddft}, some of which were performed
within TDDFT \cite{benzene-tddft}. The KS equations for an isolated
benzene molecule were solved using periodic boundary conditions (PBC)
in a tetragonal simulation cell whose base, parallel to the carbon
ring (`$xy$' plane), has a side length 30.0 \r{A}, and with height 2/3
of this; the CC and CH bond lengths were set to 1.40 \r{A} and 1.09
\AA, respectively. A simple LDA exchange-correlation functional was
adopted \cite{PZ}, whereas norm-conserving pseudopotentials from the
{\tt PWscf} table were used \cite{PP} with a PW kinetic-energy cutoff
of 60.0 Ry. The PBC calculation of dipole matrix elements was
performed as explained in Sec.~C.2 of Ref. \onlinecite{RMP}. The
absorption coefficient was obtained as: $I(\omega) \propto \omega {\rm
  Im} \chi({\omega})$, where $\chi(\omega)$ is the electric dipole
susceptibility, Eq.~(\ref{eq:susceptibility}), calculated at complex
frequencies with an imaginary part of 0.27 eV.

The convergence properties of our block-Lanczos algorithm are
displayed in Fig.~\ref{fig:benzene_full_int_convergence} where we
report the absorption spectra of benzene as calculated for light
polarized in the $xy$ plane, for different numbers of recursion
steps. We see that 2000-3000 steps are sufficient to ensure
convergence for energies up to $\approx 15~\rm ev$. Our most converged
spectrum (corresponding to 6000 recursion steps) is then compared with
that obtained using the real-time propagation method (see the
inset). The agreement between the spectra obtained with these rather
different methods is practically complete. Although a thorough
theoretical analysis of the convergence properties of our algorithm is
beyond the scope of the present letter, we would like to point out a
few facts that, we believe, will deserve further attention. We first
notice that, not unexpectedly, the convergence properties deteriorate
with increasing frequency: the lower the frequency, the better the
convergence. Second, the convergence rate of the algorithm seems to be
somewhat affected by the condition number of the Liouvillian which, in
turn, depends on the size of the PW basis set: the higher the PW
kinetic-energy cutoff, the worse the convergence
\cite{silane}. Furthermore, the numerical instabilities which are
known to plague Lanczos diagonalization algorithms \cite{Bai-and-Day}
seem to have little, if any, effect on the calculation of the
resolvent matrix elements through
Eq. (\ref{eq:continued-fraction}). As far as we can say,
Eq. (\ref{eq:continued-fraction}) can be pushed as much as needed to
reach any desired level of accuracy. Finally, the non-Hermitean
character of the Liouvillian seems to affect somewhat the efficiency
of the algorithm, as can be seen from the performance of the TDA that
we examine now.

\begin{figure}[!bt]
  \begin{center}
    \includegraphics[clip,width=80mm]{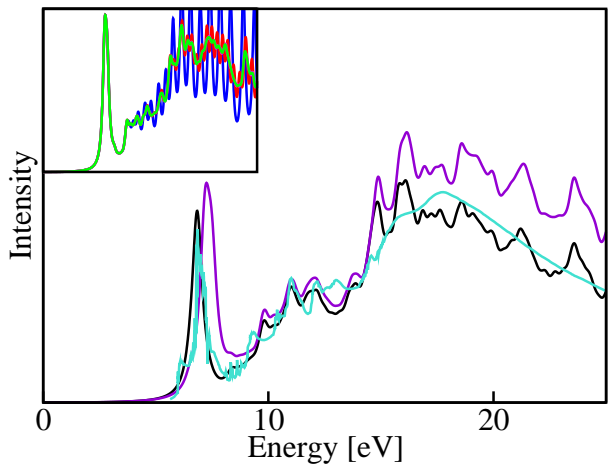}
    \caption{
      \label{fig:benzene_full,TD,expt} 
      Comparison between the converged Lanczos spectrum obtained with
      the full non-Hermitean Liouvillian (black), the Tamm-Dancoff
      approximation (red) and the experimental results (light
      blue). The inset shows the convergence of the TDA spectrum with
      respect to the number of recursion steps: 250 (blue), 500
      1000 (magenta), and 2000 (orange).
    }
  \end{center}
\end{figure}

In Fig.~\ref{fig:benzene_full,TD,expt} we compare the experimental
absorption spectrum of Ref. \cite{benzene-expt} with those obtained
from our TDDFT method, with and without use of the TDA. The agreement
between the calculated and experimental spectra is very good, as was
already known from previous TDDFT calculations for benzene
\cite{benzene-tddft}. The quality of this agreement is due to the
nature of the molecular orbitals involved in the transitions which
dominate the low-lying part of the spectrum ($\pi$, $\pi^*$, and, to a
lesser extent, $\sigma$), which are little affected by the wrong
asymptotic behavior of the ALDA XC potential. The absorption spectra
reported in Fig.~\ref{fig:benzene_full,TD,expt}, though resulting from
an average over different polarizations, are dominated by the
$(\pi^*\leftarrow \pi)\ ^1E_{1u}$ transition which is only allowed
when the light is polarized in the plane of the molecule. This
transition is mainly responsible for the first strong absorption peak
experimentally found at 6.94 eV, and predicted by TDDFT at 6.83 eV. It
is interesting to notice that the $z$ component of the spectrum
displays a weak peak at 6.55 eV, which is not visible in
Fig.~\ref{fig:benzene_full,TD,expt} (its intensity is more than 10
times smaller than the $xy$ peak), and which corresponds to a
$(\pi^*\leftarrow \sigma)\ ^1A_{2u}$ transition. In the
independent-electron approximation, this transition would have a
higher energy than $^1E_{1u}$ which corresponds to the HOMO-LUMO
gap. The red shift of the $^1A_{2u}$ transition is therefore due to
the effects of the electron-electron interaction which are
approximately accounted for in ALDA-TDDFT. The $^1A_{2u}$ transition
has never been detected directly in absorption experiments, but its
existence (as well as its location near, possibly at a lower frequency
than, the strong $^1E_{1u}$ transition) was inferred from Raman
scattering experiments \cite{raman}. The proximity of the $^1E_{1u}$
and $^1A_{2u}$, as well as the much smaller intensity of the latter,
were also confirmed by accurate coupled-cluster calculations
[\onlinecite{benzene-theory}d]. As far as we can tell, it is not
impossible that the little shoulder observed at 6.19 eV in the
absorption spectrum \cite{benzene-expt} and attributed to a
vibron-assisted $^1B_{1u}$ forbidden excitation, is actually due to a
weak $^1A_{2u}$ allowed transition.

Use of the TDA does not change much the overall appearance of the
spectrum, nor the positions of the peaks, the main difference
regarding their intensities. It is worth noticing that the convergence
of TDA calculations appears to be much faster than when using the full
non-Hermitean form of the Liouvillian (see the inset of
Fig.~\ref{fig:benzene_full,TD,expt}).

We believe that the method presented in this letter, while not
touching our still serious ignorance about the form of time-dependent
XC functionals, will open the way to a systematic study of systems
which are too large to be treated with currently available
methods. The already rather favorable numerical features of our method
will be further improved either by implementing the use of ultra-soft
pseudopotentials (which will allow reduction of both the size of
one-electron basis sets and the condition number of the Liouvillian),
and by devising optimal strategies for restarting the Lanczos chain
using approximate schemes. Work is in progress along these lines.

One of us (SB) wishes to thank Y. Saad for very useful discussions, as
well as A. Polian for hospitality at the Universit\'e Pierre et Marie
Curie, Paris, where this work was started, and R. Wentzcovitch for
hospitality at the University of Minnesota, where most of this paper
was written.

\end{document}